\documentclass[10pt,letterpaper]{article}
\usepackage{opex3}
\usepackage{graphicx}
\usepackage{dcolumn}
\usepackage{bm}
\usepackage{array}
\usepackage{hyperref}
\usepackage{amsmath}
\usepackage{multirow}
\usepackage{cite}

\begin{document}
\title{Elimination of polarization degeneracy in circularly symmetric  bianisotropic waveguides: a decoupled case}
\author{Jing Xu, Bingbing Wu, and Yuntian Chen$^*$}
 \address{School of Optical and Electronic Information, Huazhong University of Science and Technology, Wuhan, China.}
\email{$^*$yuntian@hust.edu.cn}

\begin{abstract}
Mode properties of circularly symmetric waveguides with one special type of bianisotropy  are studied using finite element approach. We find that the polarization degeneracy in circularly symmetric waveguides can be eliminated, by introducing intrinsic crossing coupling between electric and magnetic moments in the constituent units of the waveguide media.  Breaking the polarization degeneracy in high order mode groups is also confirmed numerically.  With the bianisotropic parameters chosen in this work, the $x$ and $y$-polarized modes  remain decoupled.  Typically, the $y$-polarized modes remain completely unchanged, while the $x$-polarized modes are turned into leaky modes that are lossy along propagation direction.  A perturbation model from coupled mode theory is developed to explain the  results and shows excellent agreement. Such asymmetric behavior between different polarizations might be feasible and useful for developing compact  polarizers in terahertz or  mid-infrared regime.
\end{abstract}


\section{Introduction}
In metamaterials, the most general description of material properties is bianisotropic constitutive relation, which governs material responses to the electromagnetic (EM) fields, as well as the cross terms, i.e., magnetic (electric) dipoles induced by the incident electric (magnetic) fields. Conventionally, the cross terms are hidden in the higher order spatial derivatives of the electric polarization since there is an ambiguity of splitting the induced current into polarization and magnetization in the macroscopic electromagnetics \cite{Landau1984}. However, by properly selecting the `gauge field' for the polarization and magnetization, one is able to treat electric and magnetic fields on the same footing \cite{Serdyukov2001}, thus be able to employ the duality between electric  and magnetic fields to simplify the analysis significantly. It was  found  that most of moving media are bianisotropic by Kong \cite{Kong1972,KongBook}, who invented the concept of bianisotropy to describe the electromagnetic properties of such kind of materials.

One  might expect exotic behavior of EM waves, or novel  devices may exist \cite{KhanikaevNatMaterial2013,ChenArxiv2013}, due to the extra degree of freedom from materials, i.e., the EM coupling in constituent units or the  building blocks. Indeed, extensive research activities, including experimental efforts \cite{Kriegler2010JQE,Gansel2009NatPhys,Wong2012Science} and theories \cite{Sersic2011,Pendry2004,GiessenOE2012,Giessen2014PRX,TangPRL2010},  have been carried out in  conceiving and synthesizing bianisotropic metamaterials. As a typical sub-class of bianisotropic media, chiral  media are reciprocal, and can be useful in  circular dichroism spectroscopy \cite{NarushimaPCCP2013}, and polarization control. On one hand, there are considerable efforts on the fundamental level to generate chiral light, or super-chiral light from chiral or achiral photonic structures, i.e., helix metallic structures \cite{Gansel2009NatPhys,YangzhenyuOE2013} and planar photonic lattices \cite{GiessenOE2012,DavisPRB2013}.  On the other hand, Engheta and others seek applications using chiral effects,  i.e., chirowaveguides \cite{Engheta1989OL,Pelet1989OL} and chiral fibers \cite{Kopp2013JLT}.   Chirowaveguides proposed by Engheta are realized by filling cylindrical waveguides  with isotropic chiral media, while chiral fibers are experimentally feasible with current fiber technology, i.e., by microforming glasses or decorating glass surfaces to create helical structures. Inspired by Engheta's pioneering work \cite{Pelet1989OL}, chirowaveguides are examined by many authors \cite{Li2010OL,Cao2011josab,Svedin1990IEEE,Paiva1992OL}. We note that in all the aforementioned  work, the chiral media are isotropic.  As for light, it corresponds to a macroscopic model where chiral molecules are randomly distributed in the host, while for  microwave the counterpart is the  wire helices with random orientations in the host media. However,  most of the available experiments in the field of chiral metamaterials \cite{Gansel2009NatPhys,YangzhenyuOE2013} suggest that it is very likely that the synthesized chiral media have strong bianisotropy. To the best of our knowledge, an efficient tool of analyzing bianisotropic waveguides is in lack in literature.

In this paper, we extend  the aforementioned work on chirowaveguides to the regime where the media can be bianisotropic using a finite element  approach. Particularly, we study the mode properties of bianisotropic waveguides with a special emphasis on how the guided $x$- and $y$-polarized modes are impacted by bianisotropy. In contrast to Engheta's findings that TE and  TM modes are entangled together, we found that our bianisotropic waveguides essentially function as linear polarizers. The elimination of polarization degeneracy \cite{BassentOE2002} typically refers to two effects, i.e., birefringence and different propagation lengths of the $x$- and $y$-polarized modes. The  difference between  real parts of effective modal indices ($\text{Re}(n_{eff})$) accounts for the birefringence, while different $\text{Im}(n_{eff})$ leads to different propagation lengths.

The paper is organized as follows. In Section 2, we outline the geometry under investigation and give the theoretical foundation  of how the finite element approach can be applied  for bianisotropic waveguides. In Section 3, we study the  mode dispersion and give a coupled mode model to explain our results. Finally, the paper is concluded in Section 4.

\section{Bianisotropic  waveguides and theoretical foundation of FEM modelling}
\subsection{Bianisotropic  waveguides}
\begin{figure}\centering
\includegraphics[scale=0.35]{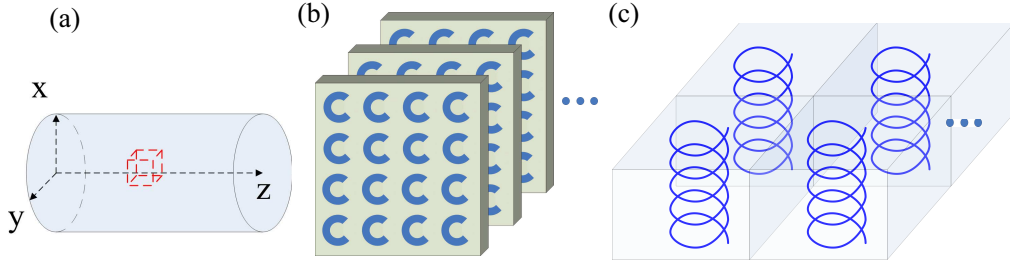}
\caption{\label{stru12}Schematic diagram of a circularly symmetric waveguide (a) which contains bianisotropic medium in the core layer by including structures with split ring resonator (SRR) arrays introducing $\chi_{12}$ (b), and structures with helix arrays introducing $\chi_{11}$ (c).}
\end{figure}

The bianisotropic waveguide studied in this work is sketched in Fig.~\ref{stru12}(a). The constitutive relation can be given as follows,
\begin{equation}\label{constitute22333}
\begin{array}{c}
\bm D= \bar{ \bm{\epsilon}} \bm E + \bar{ \bm{\chi}} _{eh} \bm H, \\
\bm B= \bar{ \bm{\mu}}   \bm H + \bar{ \bm{\chi}} _{he} \bm E,
\end{array}
\end{equation}
where $\bar {\bm \epsilon}= \epsilon _{0}\bar {\bm \epsilon}_r$, $\bar {\bm \mu}=\mu_0\bar {\bm \mu}_r$, $ \bar{ \bm{\chi}} _{he}=\sqrt{ \epsilon _{0} \mu_0}\bar{\bm{\chi}}_{he}^r$, $ \bar{ \bm{\chi}} _{eh}=\sqrt{ \epsilon _{0} \mu_0}\bar{\bm{\chi}}_{eh}^r$. Reciprocity imposes the constraints on the material parameters as given by $\bar{ \bm{\epsilon}}=\bar{ \bm{\epsilon}}^{T}$, $\bar{ \bm{\mu}} =\bar{ \bm{\mu}} ^{T}$, $\bar{ \bm{\chi}} _{eh}= -\bar{ \bm{\chi}} _{he}^{T}$. For lossless  bianisotropic media, $\bar{ \bm{\chi}} _{eh}$ is purely imaginary. Bianisotropic media in principle can be realized by aligning electrically small  magnetoelectric inclusions along particular directions. Figure~\ref{stru12}(b) and \ref{stru12}(c) show two examples of aligned dipoles from small structures, i.e. split-ring resonators (SRRs) and helix arrays,  respectively, filling a typical block (dashed region) shown in Fig.~\ref{stru12}(a). It should be noted that  effective constitutive parameters  of SRRs and helix arrays depend on many factors such as geometric shape, lattice constant, and incident polarization \cite{LiuNa_np,WuPRL2011,MingNaiBenPRB2010}. The bianisotropy  could be very dispersive as well.

In our case, we study bianisotropic waveguides operating at a single frequency. The dimensions of SRRs or helix arrays should be far smaller than the wavelength, which implies that structure details, i.e., geometric shapes and lattice constant of the meta-atoms, are averaged out, except the bianisotropic response built from the aligned meta-atoms. Under this assumption, applying a magnetic field along the $y$-direction on split ring resonators (SRRs) will induce electrical currents flowing over the SRRs along clock or anti-clock direction, which subsequently generate electric dipoles orientated along $x$-direction, as shown in  Fig.~\ref{stru12}(b). This corresponds to an off-diagonal term $\chi_{12}$ in $\bar{\bm{\chi}}_{eh}$. Figure~\ref{stru12}(c) shows a typical structure that can be used to generate the diagonal term  $\chi_{11}$ in $ \bar{ \bm{\chi}} _{eh}$.
%

\subsection{Theoretical foundation of FEM modelling}
We describe the theoretical foundation of the finite element method that is employed in this work. We take a time harmonic dependence $e^{i \omega t }$ for the EM waves throughout this paper. One can rewrite Eq.~(\ref{constitute22333}) and insert it into source-free Maxwell's equations,
\begin{equation}\label{TETMrelation3}
\begin{array}{c}
\nabla  \times \bm E = - i\omega   [\bar{ \bm{\mu}} \bar{ \bm{\chi}} _{eh} ^{-1}   \bm D+ ( \bar{ \bm{\chi}}_{he} -  \bar{ \bm{\mu}} \bar{ \bm{\chi}} _{eh} ^{-1}\bar{ \bm{\epsilon}}  ) \bm E], \\
\nabla  \times(\bar{ \bm{\chi}} _{eh} ^{-1}   \bm D- \bar{ \bm{\chi}} _{eh} ^{-1}\bar{ \bm{\epsilon}}   \bm E) = i\omega  \bm D. \\
\end{array}
\end{equation}
Combining the two equations in Eq.~(\ref{TETMrelation3}), one can calculate the eigenmodes of the bianisotropic waveguides  as given in the  following equation,
\begin{equation}\label{BianisotropicWaveguidek01}
 - \nabla  \times  \bar{ \bm{\mu}}_r^{-1} \nabla  \times \bm E(\bm r,\omega )   +    k_0 \bar{\bm{c}}_1 \nabla  \times \bm E(\bm r,\omega )  +   k_0   \nabla  \times \bar{\bm{c}}_2 \bm E(\bm r,\omega ) +k_0^2 (\bar{\bm d}_1+ \bar{\bm d}_2)\bm E(\bm r,\omega )=0,
\end{equation}
where $\bar{\bm d}_1=  \bar{ \bm{\epsilon}_r}  $, $\bar{\bm d}_2=  -\bar{ \bm{\chi}} _{eh} ^{r}\bar{ \bm{\mu}}_r^{-1}  \bar{ \bm{\chi}}_{he}^{r} $, $\bar{\bm{c}}_1= i \bar{ \bm{\chi}} _{eh} ^{r}\bar{ \bm{\mu}}_r^{-1}$, $\bar{\bm{c}}_2 =-i\bar{ \bm{\mu}}_r^{-1}  \bar{ \bm{\chi}}_{he}^{r} $, and $k_0=\omega_0\sqrt{\epsilon_0\mu_0}$.
Due to the translation symmetry of the waveguides, we can expand the modes in combination of plane waves along $z$-direction, i.e., $\bm E(\bm r)=\int\limits_{-\infty}\limits^{\infty} \bm E(x,y,\beta) e^{-i\beta z} d\beta$, where $\bm E(x,y,\beta) =\hat{x} E_x(x,y,\beta)+\hat{y} E_y(x,y,\beta)+\hat{z} E_z(x,y,\beta)$. Substituting $\bm E(\bm r)$ into Eq.~(\ref{BianisotropicWaveguidek01}) in conjunction with the test function of $\bm F(x,y,z)=\bm F(x,y,\beta_2) e^{-i\beta_2 z} $, we can get the variational form of Maxwell's equation for the waveguide modes give by  2-dimensional (2D) integration over the transverse plane ($S$),
\begin{equation}\label{2dweak}
\begin{split}
L=&-\iint\limits_{S} dxdy    {\frac{ 1} {{\bar{\bm{\mu}}_r }}  \bm {CurlE}(x,y,\beta)  \cdot  \bm {CurlF}^ *(x,y,\beta)} + \oint\limits_{\partial S}{\bm F^ *  (x,y,\beta ) \cdot [\frac{1}{{\bar{\bm{\mu}} _r }}\bm n \times \bm {CurlE}(x,y,\beta )]} dl \\& + \iint\limits_{S} dxdy  k_0[  \bar{\bm { c}}_1 \nabla  \times  + \nabla  \times \bar{\bm { c}}_2  +k_0(\bar{\bm d}_1+ \bar{\bm d}_2)]  \bm E(x,y,\beta)  \cdot  \bm F^ *(x,y,\beta),
 \end{split}
 \end{equation}
 where $\bm {CurlE}(x,y,\beta)=[\hat{x}(\frac{\partial{E_z^\beta}}{\partial{y}}+i\beta E_y^\beta)
+\hat{y}(-\frac{\partial{E_z^\beta}}{\partial{x}}-i\beta E_x^\beta)+
\hat{z}(\frac{\partial{E_y^\beta}}{\partial{x}}-\frac{\partial{E_x^\beta}}{\partial{y}})]$.
For guided modes, the fields in the transverse plane decay to zero on the boundaries, as the modelling domain is relatively large, i.e., several tens  of wavelength. Hence the term of the boundary integration in Eq.~(\ref{2dweak}) drops out. As regards to Eq.~(\ref{2dweak}), there are a few facts and implications that we intend to discuss. Firstly, Eq.~(\ref{2dweak}) is corresponding to the functional of  the reduced  wave equation,  and can be implemented in a truncated 2D computation domain, using standard finite element procedure. Secondly, Eq.~(\ref{2dweak}) can be turned into an eigenvalue problem, in which the  propagation constant $\beta$ and the field solutions  $\bm E(x,y,\beta)$ can be solved directly. Thirdly, Eq.~(\ref{2dweak}) is capable of handling waveguides consisting of bianisotropic materials, in either core or cladding layer, or both of them. In the following section, we will use this numerical tool to study a circularly symmetric waveguide, in which the material of the core layer is bianisotropic. Further details of implementation of FEM calculations can be found in \cite{ycheprb2010,ycheoe2010,Yanjosab2007,Jinbook2002}.

\section{Mode dispersion and coupled mode model}
\subsection{Dispersion}
Starting from the simplest case,  single mode cylindrical waveguides are considered in the paper. As shown in Fig.~\ref{stru12}(a), the geometry is identical to  conventional cylindrical waveguide with high material index in the core surrounded by air, except that the  core layer is bianisotropic. Explicitly, the cross term is given by
 $\bar{\bm{\chi}}_{eh}^r=-[\bar{\bm{\chi}}_{he}^r]^T=i \left(
\begin{array}{ccc}
 0 & \chi _{12} & 0 \\
0 & 0 & 0 \\
0 & 0 & 0 \\
\end{array}
\right)$, which corresponds to the scenario sketched in Fig.~\ref{stru12}(b). Rigourously speaking, bianisotropic parameters, permittivity and permeability are not completely independent from each other. For instance, $\bar{\bm{\chi}}_{eh}$ of homogenous medium as well as that of magnetoelectric point scatters suffer a upper bound \cite{Sersic2011} imposed by the magnitude of the electric and magnetic responses, i.e., $\chi_{12}\le\sqrt{\epsilon_r \mu_r}$. In this paper, we approximate the bianisotropic parameters as independent parameters from permittivity and permeability, which may hold for the media that are off structure-resonances. Without further complications, the  permittivity and permeability are approximately taken as constant scalar values.

\begin{figure}\centering
\includegraphics[scale=0.45]{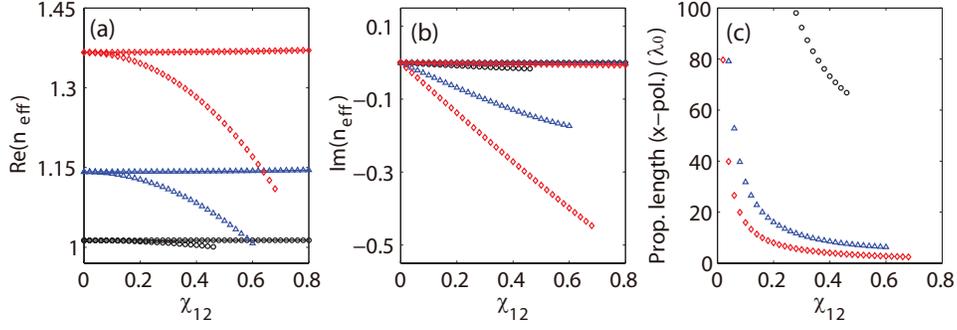}
\caption{\label{dispersion} (a) $\text{Re}(n_{eff})$, (b) $\text{Im}(n_{eff})$, (c) propagation length of $x$-polarized modes. Symbols with (without) lines refers to $y$- ($x$-) polarization modes. The color indicates different size of the radius, i.e., 0.12$\lambda_0$ (black circles), 0.16$\lambda_0$ (blue triangles) and 0.2$\lambda_0$ (red diamonds) respectively, where $\lambda_0$ is vacuum wavelength.  We use $\epsilon_{11}=\epsilon_{22}=\epsilon_{33}=4$, $\mu_{11}=\mu_{22}=\mu_{33}=1$.  The other elements in $\bar{\bm{\chi}}_{eh}^r$ is set to be  zero. }
\end{figure}

Figure~\ref{dispersion}(a) and \ref{dispersion}(b) show  $\text{Re}(n_{eff})$ and $\text{Im}(n_{eff})$ as a function of $\chi_{12}$ for three core radii. The sizes of the waveguides are chosen so that only fundamental modes with two orthogonal polarizations are supported. It can be seen that $y$-polarized modes keep almost unchanged, while $x$-polarized modes are altered significantly. In other words, the degeneracy between $x$- and $y$-polarized modes, a commonly feature of circularly symmetric waveguides, is broken. In addition, it is clear that two orthogonally polarized modes are decoupled. Such property is distinct from isotropic chiral media or chirowaveguides where $x$- and $y$-polarization are entangled together. Furthermore, for $x$-polarized modes, $\text{Re}(n_{eff})$ drops while $|\text{Im}(n_{eff})|$ increases as $\chi_{12}$ increases. This means that $x$-polarized modes become lossier as $\chi_{12}$ increases. To illustrate this, we plot the propagation lengths as a function of $\chi_{12}$ when the powers of $x$-polarized modes are attenuated by 1000 times (30dB) compared to the case of $\text{Im}(n_{eff})=0$ , as shown in Fig.~\ref{dispersion}(c). Thus high polarization extinction ratio might be realized over a propagation distance of a few wavelengths.

Figure~\ref{dispersion} also shows the different behavior of $n_{eff}$ for three different waveguides. In addition, $x$-polarized modes vanish upon certain $\chi_{12}$ values. In the following, we develop a coupled mode model to explain these phenomena.

\subsection{Coupled mode theory in bianisotropic waveguides}
To understand how the modes are impacted by the bianisotropy, it is useful to examine constitutive relation of the bianisotropic medium firstly. The three components of $\bm{D}$ and $\bm{B}$ can be written as
\begin{equation}\label{constitute}
\begin{array}{ll}
D_x=\epsilon_0(\epsilon_{11}e_x+i\chi_{12}h_y), &B_x=\mu_0\mu_{11}h_x;\\
D_y=\epsilon_0\epsilon_{22}e_y, &B_y=\mu_0(-i\chi_{12}e_x+\mu_{22}h_y);\\
D_z=\epsilon_0\epsilon_{33}e_z, &B_z=\mu_0\mu_{33}h_z.
\end{array}
\end{equation}
It is clear from Eq.~(\ref{constitute}) that $D_y$ and $B_x$ have no contribution from magnetoelectric coupling in our configuration. Therefore, $y$-polarized modes are not affected since they are dominated by $E_y$  and $H_x$. On the contrary, $x$-polarized modes are dominated by $E_x$ and $H_y$, hence they are strongly modified due to $\chi_{12}$.  In the following, we construct a coupled mode model to study how the  $x$-polarized modes evolve as $\chi_{12}$ increases.

To simplify our analysis, we use normalized  fields $[\bm e, \bm h]$, which are given by $[\bm e, \bm h]=[\bm{E}, Z_0\bm{H}]$ with $Z_0=\sqrt{\frac{\mu_0}{\epsilon_0}}$. Expressing $\bm{e}$ and $\bm{h}$ as $\bm{e}(\bm{r})=\bm{e}(x,y,\beta)e^{-i\beta z}$, $\bm{h}(\bm{r})=\bm{h}(x,y,\beta)e^{-i\beta z}$,  we reformulate the Maxwell's equation for the bianisotropic waveguide modes  as follows
\begin{subequations} \label{Maxwellwaveguide1}
\begin{align}
\nabla_t\times\bm{e}^{2d}-i\beta\bm{z}\times\bm{e}^{2d}=-ik_0(\bar{\bm{\mu}}_r\bm{h}^{2d}+\bar{\bm{\chi}}_{he}^{r}\bm{e}^{2d}),\\
\nabla_t\times\bm{h}^{2d}-i\beta\bm{z}\times\bm{h}^{2d}=ik_0(\bar{\bm{\epsilon}}_r\bm{e}^{2d}+\bar{\bm{\chi}}_{eh}^{r}\bm{h}^{2d}),
\end{align}
\end{subequations}
where $\bm{e}^{2d}=\bm{e}(x,y,\beta)$, $\bm{h}^{2d}=\bm{h}(x,y,\beta)$, $\nabla_t=\bm{x}\frac{\partial}{\partial x}+\bm{y}\frac{\partial}{\partial y}$, and $\beta$ is the propagation constant. Applying Eq.~(\ref{Maxwellwaveguide1}) to two modes with different propagation constants and fields, namely,  $\beta_0$ ($[\bm e_0^{2d}, \bm h_0^{2d}]$) and $\beta$ ($[\bm e^{2d}, \bm h^{2d}]$), we derive the formula of $\Delta\beta$ as given by
\begin{equation}\label{beta}
\Delta\beta=\frac{-i\oint(\bm{e}\times\bm{h}_0^*)\cdot\bm{n}dl+\iint {\{(k_0\bar{\bm{\mu}}_r\bm{h})\cdot\bm{h}_0^*}-(k_0\bar{\bm{\epsilon}}_r\bm{e}_0^*)\cdot\bm{e}\} dxdy}{\iint {\bm{z}\cdot(\bm{e}\times\bm{h}_0^*)} dxdy}-ik_0\Delta\chi_{12}\frac{\iint_{core}e_xh_{0y}^*dxdy}{\iint \bm{z}\cdot(\bm{e}\times\bm{h}_0^*)dxdy},
\end{equation}
where $\Delta\beta=\beta-\beta_0$, $\Delta\chi_{12}=\chi_{12}-\chi_{12,0}$, $\bm n$ the normal vector of the outer boundary.  We refer to Refs. \cite{HausJLT1987,HuangJOSAA1994,Harrington2001} for further details on the derivation. It is important to point out the fact that $x$- and $y$-polarized modes are decoupled, which is  in consistency with the dispersion relation shown in Fig.~\ref{dispersion}, and the analysis of Eq.~(\ref{constitute}). Mathematically, it translates to the fact that there is no projection between the fields in Eq.~(\ref{beta}) from the two modes picked up from two different dispersion lines of Fig.~\ref{dispersion}. This is in marked contrast to the chirowaveguides discussed by Engheta. In the decoupled case,  Eq.~(\ref{beta}) is exact and shows quantitatively how $\beta$ evolves as $\chi_{12}$ varies. As a trivial application of Eq.~(\ref{beta}) to $y$-polarized modes, we immediately find that their propagation constants keep unchanged, since $\bar{\bm{\chi}}_{he}$ ($\bar{\bm{\chi}}_{eh}$) only picks up $E_x$ ($H_y$) components, which are far smaller than $E_y$ ($H_x$). The derivation of the coupled mode theory in coupled cases are similar. However, the final form of Eq.~(\ref{beta}) shall be modified in a matrix form, whose elements are determined by the projection between different polarization modes.
\begin{figure}\centering
\includegraphics[scale=0.44]{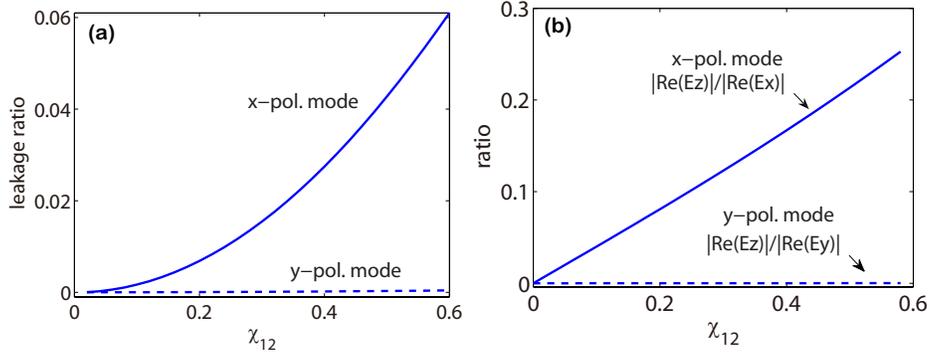}
\caption{\label{Ez_ratio3} (a) Power leakage ratio defined by $\eta$ versus $\chi_{12}$. $\eta$ is defined by Eq.~(\ref{eta}). (b) Real part of $e_z$ normalized by $e_x$ ($e_y$) for $x$-($y$-) polarized mode versus $\chi_{12}$. Radius of the waveguide in (a) and (b) is 0.2$\lambda_0$.  }
\end{figure}

\subsection{Power leakage of $x$-polarized modes}
We proceed to discuss how the $x$-polarized modes are impacted by the variation of  $\chi_{12}$. In case of no variation, i.e., $\Delta\chi_{12}=0$,  and $\bm{h}=\bm{h_0}$, $\bm{e}=\bm{e_0}$, $\Delta\beta=0$, Eq.~(\ref{beta}) is reduced to the following relation given by
\begin{equation}\label{leakage}
i\oint(\bm{e}_0^{2d}\times[\bm{h}_0^{2d}]^{*})\cdot\bm{n}dl={\iint {\{(k_0\bar{\bm{\mu}}_r\bm{h}_0^{2d})\cdot[\bm{h}_0^{2d}]^{*}}-(k_0\bar{\bm{\epsilon}}_r[\bm{e}_0^{2d}]^{*})\cdot\bm{e}_0^{2d}\} dxdy}.
\end{equation}
It is well known that for properly propagating waveguide modes, namely, perfectly guided modes, the energy stored in electric and magnetic fields shall be balanced. As evident from Eq.~(\ref{leakage}), the difference of the stored energy between electric and magnetic part equals to the leakage of power flux in the transverse plane. Indeed, Fig.~\ref{Ez_ratio3}(a) shows the leakage ratio $\eta$ as a function of $\chi_{12}$ where $\eta$ is defined as
\begin{equation}\label{eta}
\eta=\frac{{\iint {\{(k_0\bar{\bm{\mu}}_r\bm{h}_0)\cdot\bm{h}_0^*}-(k_0\bar{\bm{\epsilon}}_r\bm{e}_0^*)\cdot\bm{e}_0\} dxdy}}{{\iint {\{(k_0\bar{\bm{\mu}}_r\bm{h}_0)\cdot\bm{h}_0^*}+(k_0\bar{\bm{\epsilon}}_r\bm{e}_0^*)\cdot\bm{e}_0\} dxdy}}.
\end{equation}
It can be seen that  $\eta$ increases ( $\eta \approx$ 0  ) with $\chi_{12}$ for $x$ ($y$)-polarized modes, which is consistent with changes of $\text{Im} (n_{eff})$ of  $x$ ($y$)-polarized mode shown in Fig.~\ref{dispersion}.

The power leakage of the $x$-polarized modes can be understood by analyzing the field components, typically $z$ components, in a simple yet instructive manner. We can find   $e_z$   from transverse components of electric and magnetic fields, i.e.,
\begin{equation}
e_z=\frac{1}{-i\beta\epsilon_{33}}(\nabla_t\cdot\bar{\bm{\epsilon}}_r^{2\times2}\bm{e}_t+\nabla_t\cdot\bar{\bm{\chi}}_{eh}^{r, 2\times2}\bm{h}_t),
\end{equation}
 the real part of which increases as  $\text{Im}(\beta)$ gets large. Figure~\ref{Ez_ratio3}(b) shows the ratio of $|\frac{\text{Re}(e_z)}{\text{Re}(e_x)}|$ for $x$-polarized mode ($e_x$ dominated in this case) and $|\frac{\text{Re}(e_z)}{\text{Re}(e_y)}|$ for $y$-polarized mode ($e_y$ dominated)  at the center of the core layer as a function of $\chi_{12}$.  For $y$-polarized mode the ratio remains constantly small, while it increases continuously with $\chi_{12}$ for $x$-polarized mode. We note the $h_z$ component bears the same trend as $e_z$, which is not shown here.  The increase in $\text{Re}(e_z)$   gives rise to the power leakage in the transverse plane, as evident from the non-zero transverse components of the real part of  Poynting vector given by
\begin{equation}\label{pflow}
\begin{array}{c}
P_x=\text{Re}\{i[e_y \text{Im}(h_z)+h_y\text{Im}(e_z)]+[e_y \text{Re}(h_z)-h_y\text{Re}(e_z)]\}=e_y \text{Re}(h_z)-h_y\text{Re}(e_z),\\
P_y=\text{Re}\{-i[\text{Im}(e_z)h_x+\text{Im}(h_z)e_x]+[h_x\text{Re}(e_z)-e_x\text{Re}(h_z)]\}=h_x\text{Re}(e_z)-e_x\text{Re}(h_z),
\end{array}
\end{equation}
where  $e_x$, $e_y$, $h_x$, $h_y$ are approximately set to be real, and  $e_z$ ($h_z$) purely imaginary at $\chi_{12}=0$ but complex at $\chi_{12}\neq0$. Thus, with the assistance of Eq.~(\ref{pflow}), one can easily find that  $P_x$ and $P_y$ are not 0 for none zero $\text{Re}(e_z)$ and $\text{Re}(h_z)$ at $\chi_{12}\neq0$, which corresponds to the leakage of power in the transverse plane.

It should be noted that such power leakage also leads to difficulties in numerical calculations, as seen by the unusual cut-offs in Fig.~\ref{dispersion}. As $\chi_{12}$ increases, the fields of the $x$-polarized modes on the outer-boundary are so large that they cannot be described by the scattering boundary or perfectly electric conductor employed in our FEM modal solver. Such a problem can be solved by properly including a perfectly matched layer.

\subsection{ Perturbation}
\begin{figure}\centering
\includegraphics[scale=0.415]{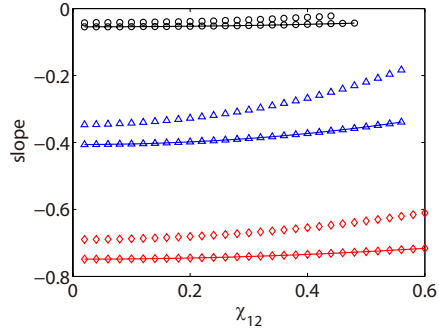}
\caption{\label{addfigure} Slope of $\text{Im}(n_{eff})$ of $x$-polarized modes calculated according to Eq.~(\ref{slope}) (symbols with lines) as well as from Fig.~\ref{dispersion}(b) (symbols without lines) versus $\chi_{12}$. Radius of the three waveguides are 0.12$\lambda_0$ (black circles), 0.16$\lambda_0$ (blue triangles) and 0.2$\lambda_0$ (red diamonds), respectively. }
\end{figure}

It is clear from Eq.~(\ref{beta}) that the first and second term on the RHS mostly contributes to the real and imaginary part variation of $\beta$, respectively. Moreover, the slope of $\text{Im}(n_{eff})$  against $\chi_{12}$ can be calculated by applying perturbation to Eq.~(\ref{beta}), assuming that the field profiles at $\beta$ remain the same as those at $\beta_0$ for tiny change of $\Delta\chi_{12}$. This approximation is the essential bit in most of the perturbation theory, and is also valid here. Following such logic and substituting  $n_{eff}=\frac{\beta}{k_0}$, we reach the following result,
\begin{equation}\label{slope}
\frac{\Delta \text{Im}(n_{eff})}{\Delta\chi_{12}}=-\frac{\iint_{core}e_{0x}h_{0y}^*dxdy}{\iint (e_{0x}h_{0y}^*-e_{0y}h_{0x}^*)dxdy}.
\end{equation}
It is clear from Eq.~(\ref{slope}) that the slopes of $y$-polarization modes are zero because $e_{0x}\approx0$ and $h_{0y}\approx0$ so that $\iint_{core}e_xh_{0y}^*dxdy=0$. Equation~(\ref{slope}) shows that the slope of $\text{Im}(n_{eff})$ is related to the magnitude of how well the in-plane components are confined in the core region. Figure~\ref{addfigure} shows the calculated slopes by extracting values from Fig.~\ref{dispersion}(b) (symbols without lines) and by our perturbation theory (symbols with lines). It indicates that Eq.~(\ref{slope}) could follow the right trend of simulation results in all cases. The general slope increases with the radius of waveguides, since the larger the radius the better the in-plane components are confined. In addition, the slopes obtained from fullwave simulation are slightly  larger than our theoretical values. The small discrepancy between simulations and theory could be attributed to the fact that the contribution of the first term of Eq.~(\ref{beta}) is neglected in our study.

\subsection{Dispersion of high order modes in bianisotropic waveguides}
Furthermore, to show the impact of bianisotropy on high order modes, the radius of the cylindrical waveguide is enlarged to form a multimode waveguide. To keep the discussion simple, the first high order vector mode group, i.e. TE$_{01}$, odd HE$_{21}$, even HE$_{21}$ and TM$_{01}$, is analyzed.

\begin{figure}\centering
\includegraphics[scale=0.60]{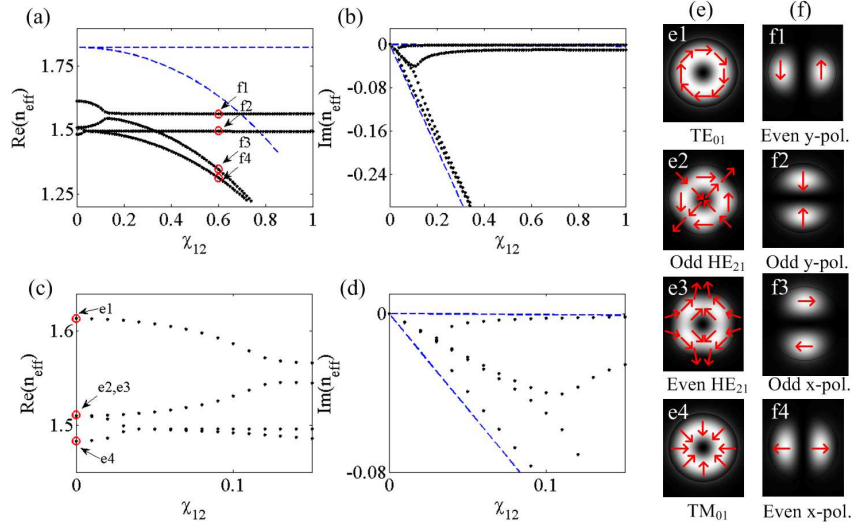}
\caption{\label{multimode} Real part (a) and imaginary part (b) of $n_{eff}$ versus $\chi_{12}$. (c) and (d) shows enlarged region of (a) and (b), respectively. (e) Mode profiles of the first high order mode group when $\chi_{12}=0$ and (f) $\chi_{12}=0.6$. Radius of the waveguide is 0.416$\lambda_0$. }
\end{figure}

Figure~\ref{multimode}(a)-\ref{multimode}(b) show the evolution of $Re(n_{eff})$ and $Im(n_{eff})$ of the first high order mode group as a function of $\chi_{12}$ (black dot lines). Figure~\ref{multimode}(c)-\ref{multimode}(d) shows enlarged mode coupling regions in Fig.~\ref{multimode}(a)-\ref{multimode}(b). The fundamental modes are shown for reference (blue dashed lines). The information in this figure is rich. First, the fundamental modes in this case behave similar to that of the single mode case. The mode profiles of the four vector modes at $\chi_{12}=0$ as marked by e1-e4 in Fig.~\ref{multimode}(c) are shown in Fig.~\ref{multimode}(e), corresponding to three different effective mode indices because odd HE$_{21}$ and even HE$_{21}$ modes are degenerate due to circular symmetry. As $\chi_{12}$ increases, mode coupling occurs between TE$_{01}$ and odd HE$_{21}$ mode as well as even HE$_{21}$ mode and TM$_{01}$ mode. This is because linear polarization (in particular $y$-polarization in the paper) is preferred by the bianisotropic medium as $\chi_{12}$ increases, the four vector modes evolve into linearly polarized modes and the degeneracy between odd HE$_{21}$ and even HE$_{21}$ mode is broken due to required mode coupling. Typical mode profiles of $x$- and $y$- polarized odd and even modes as marked by f1-f4 in Fig.~\ref{multimode}(a) are shown in Fig.~\ref{multimode}(f), corresponding to $\chi_{12}=0.6$ in this case. When linearly polarized modes are formed, $x$-polarized odd and even modes behave similar to the fundamental $x$-polarized  modes with $E_z$ components and $|\text{Im}(n_{eff})|$ increase, showing the modes turn into leaky modes. It is worth noting that the slopes of $\text{Im}(n_{eff})$ calculated by Eq.~(\ref{slope}) in this case are close to $1$, matching the values extracted from Fig.~\ref{multimode}(b). This indicates that the in-plane components, i.e., $E_x$ and $H_y$, are well confined in the core region. As shown in Fig.~\ref{multimode}(c)-\ref{multimode}(d), mode coupling behavior is complicated. However, detailed coupling behavior between high order modes is beyond the scope of this paper and are left to future work.

\section{Conclusions}
In closing, we have investigated the mode properties of circularly symmetric waveguides which contain bianisotropic media in core layers. Essentially,  we find that bianisotropy breaks the polarization degeneracy between two linear polarizations, i.e., $x$- and $y$-polarizations. The bianisotropy turns  a mode of undesired polarization ($x$-polarization in this case) from the degenerate pair into leaky mode and leaves the other one unchanged, in contrast to previous findings in chirowaveguides or chiral fibers that $x$- and $y$-polarization are entangled together. Moreover, modes with undesired polarization can be treated as effectively eliminated due to the exponential attenuation as they propagate. A simple model is  built from coupled mode theory  to study the impact from the bianisotropy, on the mode properties. A perturbation is further applied to account for the variation of imaginary part of the effective mode with respect to the chirality parameter and shows good agreement with numerical results. It is worthwhile to point out that the effective elimination of undesired polarization is general, which  also occurs in waveguides with non-circular cross-sections, and operates in multimode regime.

As an outlook, we envisage that the bianisotropic waveguides might be useful in terahertz or  middle infrared regime. Compact and integrated  polarizers is challenging but essential for devices and systems operating with a single polarization \cite{Burns_1984,Wang_oe_2007,Doerr_2010}. The bianisotropic waveguides proposed here may provide potentials for integrated linear polarizer with high polarization selectivity over compact interaction length. Regarding the realizability of bianisotropic waveguide, there are three relevant dimensions: (1) the feature size  compatible with  current fabrication technology,  (2) the size of the meta-atoms, (3) the cross-section of waveguide. To respect the effective constitutive parameter used here, the size of meta-atoms should be less than 1/10$\lambda_0$.  Therefore, fabrication of optical bianisotropic waveguides is challenging. However, it may be possible to explore  direct laser writing (DLW) combined with STED technique \cite{Wegener2014DLW} for operation at middle infrared or longer wavelengths. As the size of SRRs or helix arrays approaches half wavelength along the propagation direction, the waveguides would behave similar to 1D photonic crystals, in which the cross-section of the waveguide shall be large enough to contain at least one meta-atom. In this case, we believe it is possible to observe similar  effect in the pass band. In general, engineering bianisotropic waveguides, or bianisotropic metamaterials, may lead to  the new possibilities of manipulating  light propagation, emission and absorption, and interesting  phenomenon  such as one way transportation of photons.

\section*{Acknowledgment}
This work was supported by National Natural Science Foundation of China (Grant No. 61405067 and 61405066) and Foundation for Innovative Research Groups of the Natural Science Foundation of Hubei Province (Grant No. 2014CFA004).


\begin{thebibliography}{10}
\bibitem{Landau1984}   L. D. Landau and E. M. Lifshitz, {\it Electrodynamics of Continous Media} (Pergamon, 1960).


\bibitem{Serdyukov2001}  A. N. Serdyukov, I. V. Semchenko, S. A. Tretyakov, and A. Sihvola, {\it Electromagnetics of Bi-anisotropic Materials: Theory and Applications} (Gordon and Breach Science, 2001).


\bibitem{Kong1972}  J. A. Kong,  ``Theorems of bianisotropic media,'' Proc. IEEE {\bf 60,} 1036-1046 (1972).

\bibitem{KongBook}   J. A. Kong, {\it Electromagnetic Wave Theory} (EMW Publishing, 2008).


\bibitem{KhanikaevNatMaterial2013}  A. B. Khanikaev, S. H. Mousavi, W.-K. Tse, M. Kargarian, A. H. MacDonald, and G. Shvets,  ``Photonic topological insulators,'' Nat. Mater. {\bf 12}, 233-239 (2013).


\bibitem{ChenArxiv2013}  W. J.  Chen, S. J. Jiang, X. D. Chen, B. C. Zhu, L. Zhou, J. W. Dong, and C. T. Chan,  ``Experimental realization of photonic topological insulator in a uniaxial metacrystal waveguide,'' Nat Commun. {\bf 5,} 5782 (2014).


\bibitem{Kriegler2010JQE} C.  \'{E} Kriegler, M.  S.  Rill, S. Linden, and M. Wegener, ``Bianisotropic photonic metamaterials,'' IEEE J. Sel. Top. Quant. {\bf 16}, 367-375 (2010).


\bibitem{Gansel2009NatPhys} J. K. Gansel, M.  Thiel, M. S. Rill,  M. Decker, K. Bade, V. Saile, G. von Freymann, S.  Linden, and M.  Wegener,  ``Gold helix photonic metamaterial as broadband circular polarizer,'' Science {\bf 325}, 1513-1515 (2009).


\bibitem{Wong2012Science} G. K. Wong, M. S. Kang, H. W. Lee, F. Biancalana, C. Conti, T. Weiss, P. St. J. Russell, ``Excitation of orbital angular momentum resonances in helically twisted photonic crystal fiber,'' Science {\bf 337}, 446-449 (2012).




\bibitem{Sersic2011} I. Sersic, C. Tuambilangana, T. Kampfrath, and A. F. Koenderink,  ``Magnetoelectric point scattering theory for metamaterial scatterers,'' \prb {\bf 83,} 245102 (2011).


\bibitem{Pendry2004} J. B. Pendry, ``A chiral route to negative refraction,'' Science {\bf 306}, 1353-1355 (2004).

\bibitem{GiessenOE2012} M. Sch\"{a}ferling, X. Yin, and H. Giessen, ``Formation of chiral fields in a symmetric environment,'' \opex {\bf 20,} 26326-26336 (2012).


\bibitem{Giessen2014PRX} M. Sch\"{a}ferling, D. Dregely, M. Hentschel, and H. Giessen, ``Tailoring enhanced optical chirality: design principles for chiral plasmonic nanostructures,'' Phys. Rev. X {\bf 2}, 031010 (2012).

\bibitem{TangPRL2010} Y. Tang and A. E. Cohen, ``Optical chirality and its interaction with matter,'' \prl {\bf 104,} 163901 (2010).


\bibitem{NarushimaPCCP2013} T. Narushima and H. Okamoto, ``Circular dichroism nano-imaging of two-dimensional chiral metal nanostructures,'' Phys. Chem. Chem. Phys. {\bf 15}, 13805-13809 (2013).

\bibitem{YangzhenyuOE2013} L. Wu, Z. Yang, Y. Cheng, Z. Lu, P. Zhang, M. Zhao, R. Gong, X. Yuan, Y. Zheng, and J. Duan, ``Electromagnetic manifestation of chirality in layer-by-layer chiral metamaterials,'' \opex {\bf 21,} 5239-5246 (2013).


\bibitem{DavisPRB2013} T. J. Davis and E. Hendry, ``Superchiral electromagnetic fields created by surface plasmons in nonchiral metallic nanostructures,'' \prb {\bf 87,} 085405 (2013).


\bibitem{Engheta1989OL} N. Engheta and P. Pelet,  ``Modes in chirowaveguides,'' \ol {\bf 14}, 593-595 (1989).


\bibitem{Pelet1989OL} P. Pelet and N. Engheta,  ``The theory of chirowaveguides,'' IEEE Trans. Antennas Propag. {\bf 38}, 90-98 (1990).



\bibitem{Kopp2013JLT} V. I. Kopp, J. Park, M. Wlodawski, J Singer, D. Neugroschl, and A. Z. Genack,  ``Chiral fibers: microformed optical waveguides for polarization control, sensing, coupling, amplification, and switching,''  J. Lightw. Technol. {\bf 32}, 605-613 (2013).


\bibitem{Li2010OL} J. Li,  Q. Su, and Y. Cao,  ``Circularly polarized guided modes in dielectrically chiral photonic crystal fiber,''  \ol {\bf 35}, 2720-2722 (2010).


\bibitem{Cao2011josab}  Y. Cao,  J. Li,  and  Q. Su,``Guided modes in chiral fiber,''  \josab {\bf 28}, 319-324 (2011).


\bibitem{Svedin1990IEEE}  J. A. M. Svedin,``Propagation analysis of chirowaveguides using the finite-element method,''  IEEE Trans. Microw. Theory Techn. {\bf 38}, 1488-1496 (1990).


\bibitem{Paiva1992OL} C. R. Paiva, A. L. Topa, and A. M.  Barbosa,  ``Semileaky waves in dielectric chirowaveguides,'' \ol {\bf 17}, 1670-1672 (1992).



\bibitem{BassentOE2002} I. M. Bassett and A. Argyros, ``Elimination of polarization degeneracy in round waveguides,'' \opex {\bf 10}, 1342-1346 (2002).


\bibitem{LiuNa_np} N. Liu, H. Liu, S. Zhu, and H. Giessen, ``Stereometamaterials,'' Nat. Photonics {\bf 3}, 157-162 (2009).

\bibitem{WuPRL2011} C. Wu, H. Li, X. Yu, F. Li, H. Chen, and C. T. Chan, ``Metallic helix array as a broadband wave plate,'' \prl {\bf 107}, 177401(2011).

\bibitem{MingNaiBenPRB2010} X. Xiong, W.-H. Sun, Y.-J. Bao, M. Wang, R.-W. Peng, C. Sun, X. Lu, J. Shao, Z.-F. Li, and N.-B. Ming, ``Construction of a chiral metamaterial with a U-shaped resonator assembly,'' \prb {\bf 81}, 075119 (2010).


\bibitem{ycheprb2010} Y. Chen, T. R. Nielsen, N. Gregersen, P. Lodahl, and J. M\o rk, ``Finite-element modeling of spontaneous emission of a quantum emitter at nanoscale proximity to plasmonic waveguides,'' \prb {\bf 81}, 125431 (2010).


\bibitem{ycheoe2010} Y. Chen, N. Gregersen, T. R. Nielsen, J. M\o rk, and P. Lodahl, ``Spontaneous decay of a single quantum dot coupled to a metallic slot waveguide in the presence of leaky plasmonic modes,'' \opex {\bf 18}, 12489-12498 (2010).


\bibitem{Yanjosab2007} M. Yan and M. Qiu, ``Guided plasmon polariton at 2D metal corners,'' J. Opt. Soc. Am. B. {\bf 24}, 2333-2342 (2007).


\bibitem{Jinbook2002} J. M. Jin, {\it The Finite Element Method in Electrodynamics, 2nd} (Wiley, 2002).

\bibitem{HausJLT1987} H. A. Haus, W. P. Huang, S. Kawakami, and N. A. Whitaker, ``Coupled-mode theory for optical waveguides,'' J. Lightwave Technol. {\bf 5,} 16-23 (1987).


\bibitem{HuangJOSAA1994} W.-P. Huang, ``Coupled-mode theory for optical waveguides: an overview,'' J. Opt. Soc. Am. A {\bf 11,} 963-983 (1994).


\bibitem{Harrington2001} R. F. Harrington, {\it Time-harmonic Electromagnetic Fields, 2nd} (Wiley-IEEE, 2001).

\bibitem{Burns_1984} W. K. Burns, R. P. Moeller, C. A. Villarruel, and M. Abebe, ``All-fiber gyroscope with polarization-holding fiber,'' \ol {\bf 9}, 570-572 (1984).



\bibitem{Wang_oe_2007} Q. Wang and J. Yao, ``A high speed 2$\times$2 electro-optic switch using a polarization modulator,'' \opex {\bf 15}, 16500-16505 (2007).

\bibitem{Doerr_2010} C. R. Doerr, P. J. Winzer, Y.-K. Chen, S. Chandrasekhar, M. S. Rasras, L. Chen, T.-Y. Liow, K.-W. Ang, and G.-Q. Lo, ``Monolithic polarization and phase diversity coherent receiver in silicon,'' J. Lightwave Technol. {\bf 28}, 520-525 (2010).

\bibitem{Wegener2014DLW} J. Fischer, J. B. Mueller, A. S. Quick, J. Kaschke, C. Barner-Kowollik, and M. Wegener, ``Exploring the mechanisms in STED-enhanced direct laser writing,'' Adv. Optical Mater. doi: 10.1002/adom.201400413 (2014).






\end{thebibliography}
\end{document}